\documentclass[11pt]{article}

\usepackage{natbib}        
\usepackage[margin=1in]{geometry}
\usepackage{times}
\usepackage[utf8]{inputenc} 
\usepackage[T1]{fontenc}    
\usepackage{hyperref}       
\usepackage{url}            
\usepackage{booktabs}       
\usepackage{amsfonts}       
\usepackage{nicefrac}       
\usepackage{microtype}      
\usepackage{xcolor}         
\usepackage{amsmath}
\usepackage{graphicx}
\usepackage{enumitem}
\usepackage{float}

\usepackage{amsmath,amsfonts,bm}

\def\eqref#1{equation~\ref{#1}}

\def\1{\bm{1}}

\def\vm{{\bm{m}}}

\def\vp{{\bm{p}}}

\def\vs{{\bm{s}}}

\def\vx{{\bm{x}}}
\def\vy{{\bm{y}}}
\def\vz{{\bm{z}}}

\def\mL{{\bm{L}}}
\def\mM{{\bm{M}}}

\def\mW{{\bm{W}}}

\DeclareMathAlphabet{\mathsfit}{\encodingdefault}{\sfdefault}{m}{sl}
\SetMathAlphabet{\mathsfit}{bold}{\encodingdefault}{\sfdefault}{bx}{n}

\newcommand{\R}{\mathbb{R}}

\title{Estimating Brain Activity with High Spatial and Temporal Resolution using a Naturalistic MEG-fMRI Encoding Model}
\author{
  \textbf{Beige Jerry Jin}\\
  Carnegie Mellon University\\
  \texttt{jerryjin@andrew.cmu.edu} \\
  \and
  \textbf{Leila Wehbe}\\
  Carnegie Mellon University \\
  \texttt{lwehbe@cmu.edu} \\
}
\date{}

\begin{document}

\maketitle

\begin{abstract}
Current non-invasive neuroimaging techniques trade off between spatial resolution and temporal resolution. While magnetoencephalography (MEG) can capture rapid neural dynamics and functional magnetic resonance imaging (fMRI) can spatially localize brain activity, a unified picture that preserves both high resolutions remains an unsolved challenge with existing source localization or MEG-fMRI fusion methods, especially for single-trial naturalistic data.  We collected whole-head MEG when subjects listened passively to more than seven hours of narrative stories, using the same stimuli in an open fMRI dataset \citep{LeBelEtAl_2023}. We developed a transformer-based encoding model that combines the MEG and fMRI from these two naturalistic speech comprehension experiments to estimate latent cortical source responses with high spatiotemporal resolution. Our model is trained to predict MEG and fMRI from multiple subjects simultaneously, with a latent layer that represents our estimates of reconstructed cortical sources. Our model predicts MEG better than the common standard of single-modality encoding models, and it also yields source estimates with higher spatial and temporal fidelity than classic minimum-norm solutions in simulation experiments. We validated the estimated latent sources by showing its strong generalizability across unseen subjects and modalities. Estimated activity in our source space predict electrocorticography (ECoG) better than an ECoG-trained encoding model in an entirely new dataset. By integrating the power of large naturalistic experiments, MEG, fMRI, and encoding models, we propose a practical route towards millisecond‑and‑millimeter brain mapping.
\end{abstract}

\begin{figure}
  \centering
  \includegraphics[width=0.8\linewidth]{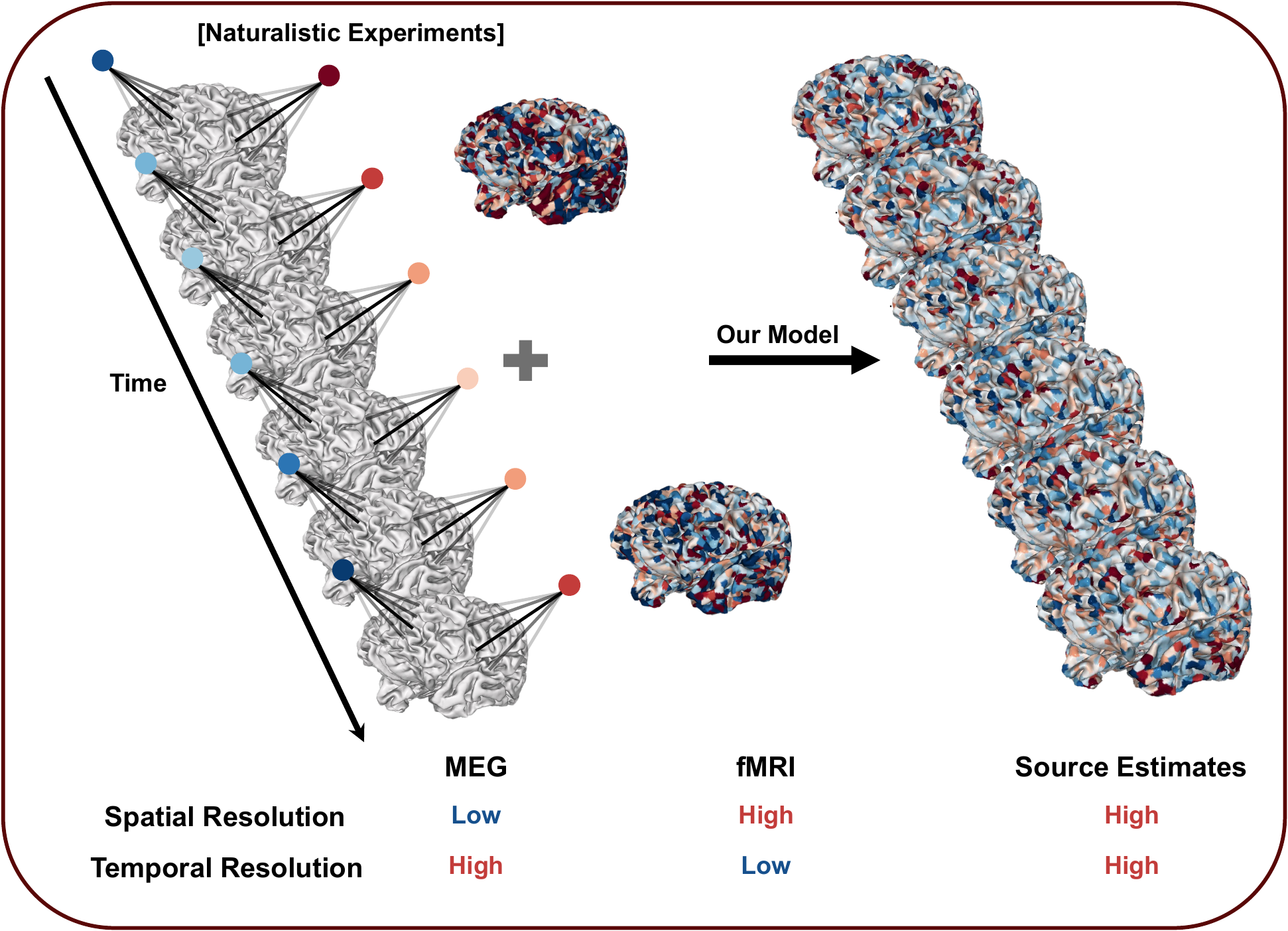}
  \caption{\textbf{Integration of MEG and fMRI.} Our work integrates the millisecond-level temporal precision of MEG with the millimeter-scale spatial specificity of fMRI to reconstruct cortical source activity at a high spatiotemporal resolution in naturalistic experiments.}
  \label{fig:teaser}
\end{figure}

\section{Introduction}
\label{sec:intro}

Non-invasive neuroimaging are central to cognitive neuroscience, yet each modality remains constrained by a fundamental trade-off between spatial and temporal resolution. Magnetoencephalography (MEG), for instance, which is sensitive to the magnetic field induced by postsynaptic current in groups of spatially aligned neurons, offers millisecond-scale temporal precision but suffers from poor spatial detail. Conversely, the blood-oxygen-level dependent (BOLD) signal measured by functional magnetic resonance imaging (fMRI) provides millimeter-scale spatial maps but reflects a sluggish hemodynamic response that integrates neural activity over seconds \citep{HallEtAl_2014} (Figure~\ref{fig:teaser}). Bridging their complementary strengths to obtain a unified, high spatiotemporal resolution view of neural source activity is critical for understanding complex processes such as speech comprehension, which recruits multiple subprocesses unfolding on the order of milliseconds across distributed cortical networks.

However, effectively integrating MEG and fMRI remains an unsolved challenge, particularly for single-trial naturalistic data. Conventional source localization methods often use fMRI data to constrain the mathematically ill-posed MEG inverse problem. For example, fMRI activation maps can serve as spatial priors in minimum-norm estimation (MNE) to improve source localization \citep{Liu_1998,Dale_2000,Suzuki_2021,Moradi_2024}. While commonly used for simple, event-related designs, these approaches ignore the stimuli and are often ineffective in recovering the continuously evolving neural dynamics in more naturalistic settings. A more promising approach, called \textit{neurogenerative modeling}, aims to directly model cortical sources and then generates MEG and fMRI signals forward \citep{Huster_2012}. Yet, prominent frameworks like The Virtual Brain often rely on complex biophysical models and anatomical connectomes, which requires much prior knowledge and makes parameter fitting inefficient. Furthermore, their focus has largely been on resting-state dynamics rather than encoding responses to external stimuli, limiting their applicability to task-based naturalistic experiments \citep{Ritter_2013,Patow_2024,Hashemi_2025}.


In this work, we introduce a novel framework that combines the strength of the neurogenerative approach with deep learning-based encoding model paradigms. We propose a novel encoding model that predicts MEG and fMRI signals for multiple subjects as a function of stimulus features, constrained by the requirement that both modalities originate from the same source estimates in a latent source space. This is achieved by incorporating anatomical information and biophysical forward models for MEG and fMRI. Thus, we effectively estimate the source activity that is high-resolution in both time and space. Crucially, the resulting source estimates are intended to be approximations that are physiologically faithful to the true brain signals, rather than a transformation of them. Note that the ground-truth neural activity at this resolution is inaccessible given current non-invasive techniques. Instead, we validate our estimates by showing that they generalize across experiments and subjects to predict invasive data recorded with electrocorticography (ECoG) from epileptic patients. Although ECoG provides a very partial coverage of the brain, each electrode provides signals that are highly resolved in time and space, and is thus an ideal test bed for our estimated brain space. We find that our model can produce powerful predictions of ECoG signal, outperforming models trained directly on ECoG data. This validates the promise of our approach at faithfully recovering the underlying brain source activity. 


\section{Methods}

\subsection{Source Space}

As a standard practice in source estimation, we first define \textit{source spaces} that specify the location of possible neural sources. For each subject, we construct a subject-specific source space according to their structural MRI scan with an octahedron-based subsampling method from \texttt{MNE-Python} \citep{Gramfort_2014}. This process yields a set of equally spaced sources on the cortical surface\footnote{This discretization is a standard step in source localization as the exact location of brain sources cannot be precisely pinpointed and has to be hypothesized as a mesh.}. We also define a source space on the ``fsaverage'' brain, a standard ``average'' brain template in \texttt{FreeSurfer} \citep{Fischl_2012}, using the same procedure. Furthermore, each source is modeled as an \textit{equivalent current dipole}, a common approximation for the net postsynaptic currents generated by a small group of spatially aligned and synchronously firing neurons. Thus, \textit{source estimates} refer to the estimated amplitudes of these dipoles over time. We additionally assume that all dipoles are oriented perpendicularly to the brain surface. 

With these anatomically-derived source spaces, we compute two matrices. Using \texttt{MNE-Python}, we calculate the source morphing matrix $\mM^{\text{S}_i}$, which transforms the source estimates from the ``fsaverage'' source space to subject $\text{S}_i$'s subject-specific source space. If $\text{S}_i$ has MEG recordings, we also compute the lead-field matrix $\mL^{\text{S}_i}$, which maps $\text{S}_i$'s source estimates to MEG sensor signals according to Maxwell's equations.

\subsection{Model Architecture}
\label{sec:model}

\begin{figure}
  \centering
  \includegraphics[width=\linewidth]{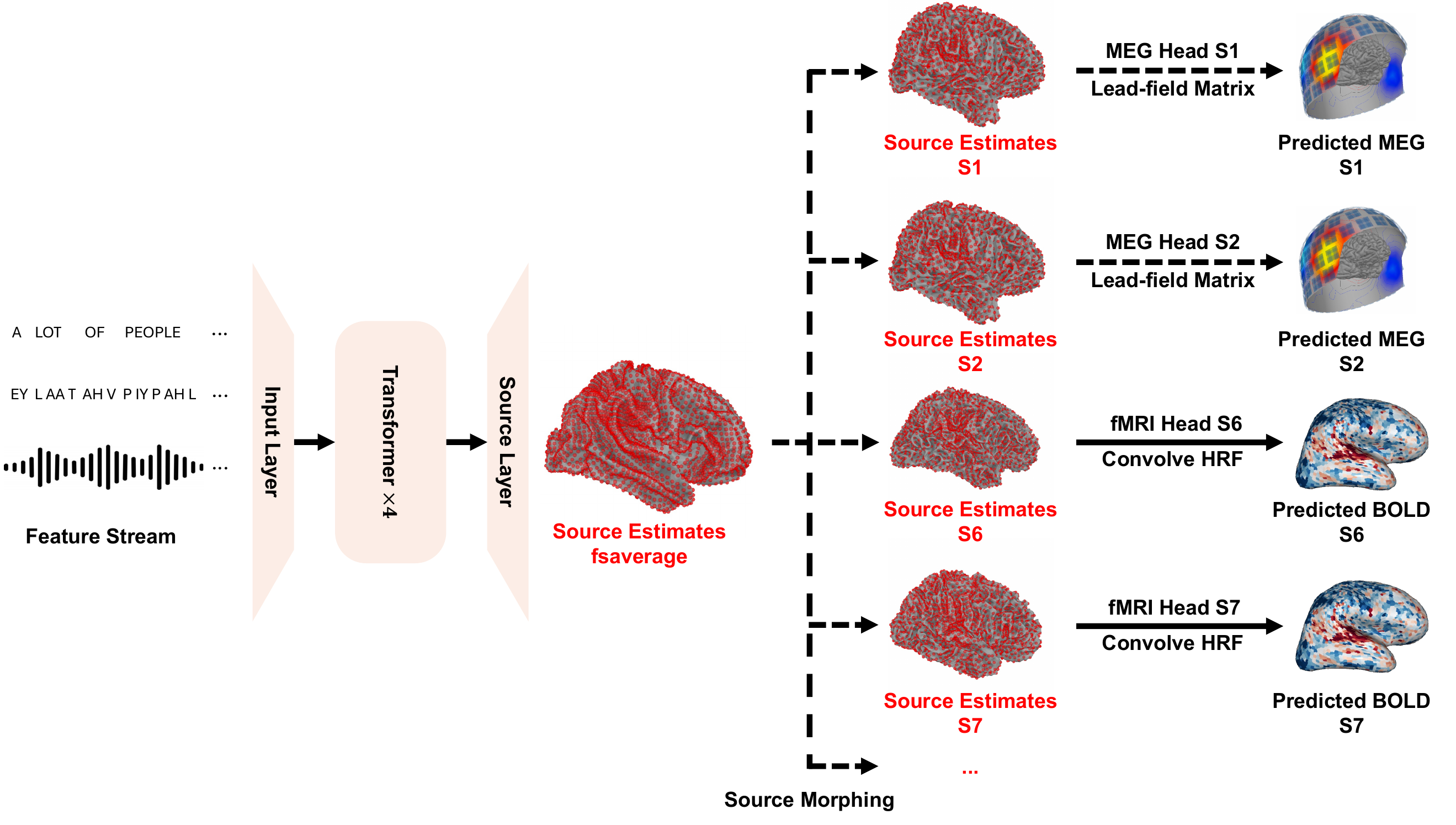}
  \vspace{-15pt}
  \caption{\textbf{Architecture of the MEG-fMRI encoding model.} Feature streams enter the network through the input layer and traverse four transformer layers before being projected into the ``fsaverage'' source space by the source layer. The source estimates in the ``fsaverage'' source space is then transformed into subject-specific source estimates by the source morphing matrix. The MEG head predicts sensor signals by multiplying the source estimates with the lead-field matrix. The fMRI head predicts BOLD responses by convolving the downsampled envelope of the source estimates with a learnable hemodynamic response function (HRF) kernel. The MEG and fMRI of multiple subjects (e.g., S1, S2, ...) are predicted simultaneously. Under the joint constraints of MEG and fMRI from multiple subjects, our model recovers the source estimates with high spatiotemporal resolution.  Dashed arrows indicate steps that are pre-computed and not learnable.}
  \label{fig:model_structure}
\end{figure}

We build a transformer-based encoding model to predict MEG and fMRI simultaneously for multiple subjects from stimulus input and source estimates, as schematized in Figure~\ref{fig:model_structure}.

\paragraph{Input Layer}
We use three concatenated feature streams to represent the naturalistic stories and serve as model input. The first feature space consists of a 768-dimensional contextual word embedding, obtained from the seventh hidden layer of GPT-2 \citep{Radford_2019} with a context window of 20 tokens. The second feature space is a phoneme feature space consisting of 44-dimensional one-hot vectors, where each dimension represents a phoneme in the CMU Pronouncing Dictionary or a non-speech sound \citep{LeBelEtAl_2023}. The third feature space is a 40-dimensional space of mel-spectrograms spanning 0-10\,kHz, representing the perceived audio sound. These yield feature vectors $\vx_t \in \R^{d_{\text{stim}}}$ with $d_{\text{stim}}=852$, sampled at 50\,Hz. The contextual word embeddings are repeated for the entirety of the word duration, and the phoneme embedding is repeated for the entirety of the phoneme duration. Then, the feature vectors go through the linear input layer before entering the transformer: $\vz_t^{\text{in}} = \mW^{\text{in}}\vx_t + \mathrm{bias}$, where $\vz_t^{\text{in}} \in \R^{d_{\text{model}}}$, $\mW^{\text{in}}\in\R^{d_{\text{model}}\times d_{\text{stim}}}$ and $d_{\text{model}}=256$.

\paragraph{Transformer Encoder}
To capture the dependency between features and feature-dependent latency in neural responses, we use four standard transformer encoder layers with two heads, feed-forward size = 512, and dropout = 0.2. Each attention block uses a causal sliding window of 500 tokens, so that the transformer could make use of the preceding 10\,s of the stimulus features. We add learnable positional embeddings to the keys within each attention block, so that the transformer can learn feature-dependent neural latency. 

\paragraph{Source Layer and Source Morphing}
The output of the transformer $\vz_t^{\text{out}} \in \R^{d_{\text{model}}}$ is then projected to the source space through the linear source layer: $\vs_{t} = \mW^{\text{src}}\vz_t^{\text{out}} + \mathrm{bias}$, where $\vs_t \in \R^{d_{\text{src}}}$, $\mW^{\text{src}}\in\R^{d_{\text{src}}\times d_{\text{model}}}$ and $d_{\text{src}}=8196$. The resulting $\vs_{t}$ would represent the \textit{source estimates} at time $t$ in the ``fsaverage'' source space. To obtain subject-specific predictions, we then use the pre-computed source morphing matrix $\mM^{\text{S}_i} \in\R^{d_{\text{src}} \times d_{\text{src}}}$ to transform $\vs_{t}$ into $\text{S}_i$'s source estimates $\vs^{\text{S}_i}_{t}$: $\vs^{\text{S}_i}_{t} = \mM^{\text{S}_i} \vs_{t}$.

\paragraph{MEG Head}
We get MEG sensor-wise prediction for $\text{S}_i$ by $\hat{\vm}^{\text{S}_i}_t = \mL^{\text{S}_i}\,\vs^{\text{S}_i}_{t}$, where $\hat{\vm}^{\text{S}_i}_t \in \R^{d_{\text{MEG}}}$ and $\mL\in\R^{d_{\text{MEG}} \times d_{\text{src}}}$ is the pre-computed lead-field matrix for $\text{S}_i$. Importantly, $\mL^{\text{S}_i}$ contains subject-specific anatomical information and guarantees that $\vs^{\text{S}_i}_{t}$ represents the source estimates in our defined source space. This, together with $\mM^{\text{S}_i}$, in turn guarantees that $\vs_{t}$ represents the source estimates in the ``fsaverage'' source space.

\paragraph{fMRI Head} 
We calculate the envelope of source estimates as $\vp^{\text{S}_i}_{t} = |\vs^{\text{S}_i}_{t} + j\,\mathcal{H}(\vs^{\text{S}_i}_{t})|$ where $\mathcal{H}$ is the Hilbert transform and $j$ is the imaginary unit. Then, we downsample $\vp^{\text{S}_i}_{t}$ and convolve it with a double-gamma hemodynamic response function (HRF; whose parameters are learnable) to yield source-level BOLD predictions $\hat{\vy}^{\text{S}_i}_{\tau} \in \R^{d_{\text{src}}}$ for $\text{S}_i$. Note that $\tau$ corresponds to the slower sampling rate of fMRI.

\subsection{Stimuli}

27 stories of ``Moth Radio Hour'' stories serve as stimuli, as in \citet{LeBelEtAl_2023}. Each story has length around 10-15 minutes. These stories are partitioned into 21 training stories, one validation story, four test stories, and one anchor story.

\subsection{MEG Data} 
\label{sec:meg}

We recorded whole-head MEG from five subjects (S1--S5) as they passively listened to the stories. The validation and test stories were presented twice and the anchor story five times, and the repetitions were averaged.  Data were acquired on a MEGIN TRIUX scanner ($d_{\text{MEG}}=306$, 204 planar gradiometers, 102 magnetometers, 102 triple-sensor locations) at 1\,kHz. The following preprocessing steps are performed with \texttt{MNE-Python} \citep{Gramfort_2013}: (1) temporal Signal Space Separation (tSSS) \citep{TauluSimola_2006}; (2) 1–150 Hz band-pass and 60/120 Hz notch filters; (3) independent component analysis (ICA) \citep{Hyvarinen_1999} removal of ocular, cardiac, and audio signal artifacts; (4) downsampling to 50 Hz. We also collected an anatomical T1 scan for each subject, which we use to reconstruct the subject's cortical surface using \texttt{FreeSurfer} \citep{Fischl_2012}.

\subsection{fMRI Data} 
\label{sec:fmri}

We use an open fMRI dataset \citep{LeBelEtAl_2023} where eight subjects (S6--S13) underwent 3\,T fMRI scans while passively listening to the stories. The reconstructed cortical surface of each subject based on the anatomical T1 scan and \texttt{FreeSurfer} \citep{Fischl_2012} is also provided. We project voxel-level data to the cortical surface using \texttt{pycortex} \citep{Gao_2015}, and then average the nearest vertices of each source to get the source-level BOLD signals.

\subsection{Model Training}
\label{sec:train}

For training efficiency and memory usage, we divide training stories into pieces of 40 seconds and train with a batch size of 8. We use the last 20 seconds to calculate loss to ensure model's full exposure to past information. Training loss $L$ is defined as a weighed sum of MEG correlation loss (weighted by the repeatability of each sensor), fMRI correlation loss (weighted by the repeatability of each source), and smoothness loss $L_\text{smooth}$:
$$
L = \alpha_1 \left(1 - \text{corr}(\vm_t, \hat{\vm}_t) \right) 
+ \alpha_2 \left(1 - \text{corr}(\vy_{\tau}, \hat{\vy}_{\tau}) \right) 
+ \alpha_3 L_\text{smooth}
$$
where $\vm_t$ and $\vy_{\tau}$ denote the MEG and fMRI data\footnote{For more effective training, we denoise the fMRI data for training stories using the predicted BOLD signals from the fMRI linear encoding model in Section~\ref{sec:pred}. Note that we still use the non-denoised fMRI data for validation and evaluation.} respectively, and $L_\text{smooth}$ is the mean squared difference between $s_t$ and $s_{t-1}$. We set $\alpha_1 = 1$ and $\alpha_3 = 0.0001$ throughout the training. For the first 20 epochs, we set $\alpha_2 = 0$ to let the model train only on MEG. For later epochs, we set $\alpha_2 = 1$ to let the model train on MEG and fMRI jointly with equal importance. This MEG-first curriculum enables our model to quickly capture the complex temporal dynamics under the guidance of MEG, and then refine the spatial pattern under the constraint of fMRI. We validate our model on the validation story and perform early stopping according to the validation loss, which has the same format as $L$.

\section{Predictive Performance}
\label{sec:pred}

We compare our model's predictive performance for MEG and fMRI against single-subject, single-modality linear encoding models.  Since these models are not constrained by the other subject, the shared source space, or the other modality, they are not simple baselines to be surpassed, but rather a high-performance benchmark representing the upper limit (``ceiling'') of what a linear model can achieve. \textbf{MEG Ridge Ceiling}: We shift the feature streams to represent the stimuli delayed by 0, 20, ..., 600\,ms, and then use these 31 stimulus embedding matrices to train a stacked ridge regression model for each MEG sensor \citep{LinEtAl_2024}. \textbf{fMRI Ridge Ceiling}: We concatenate features delayed by 2, 4, 6, and 8\,s and fit a ridge regression for the source-level BOLD signals. This is the same model used for denoising training fMRI data (see Section~\ref{sec:fmri}).

\begin{figure}
  \centering
  \includegraphics[width=0.9\linewidth]{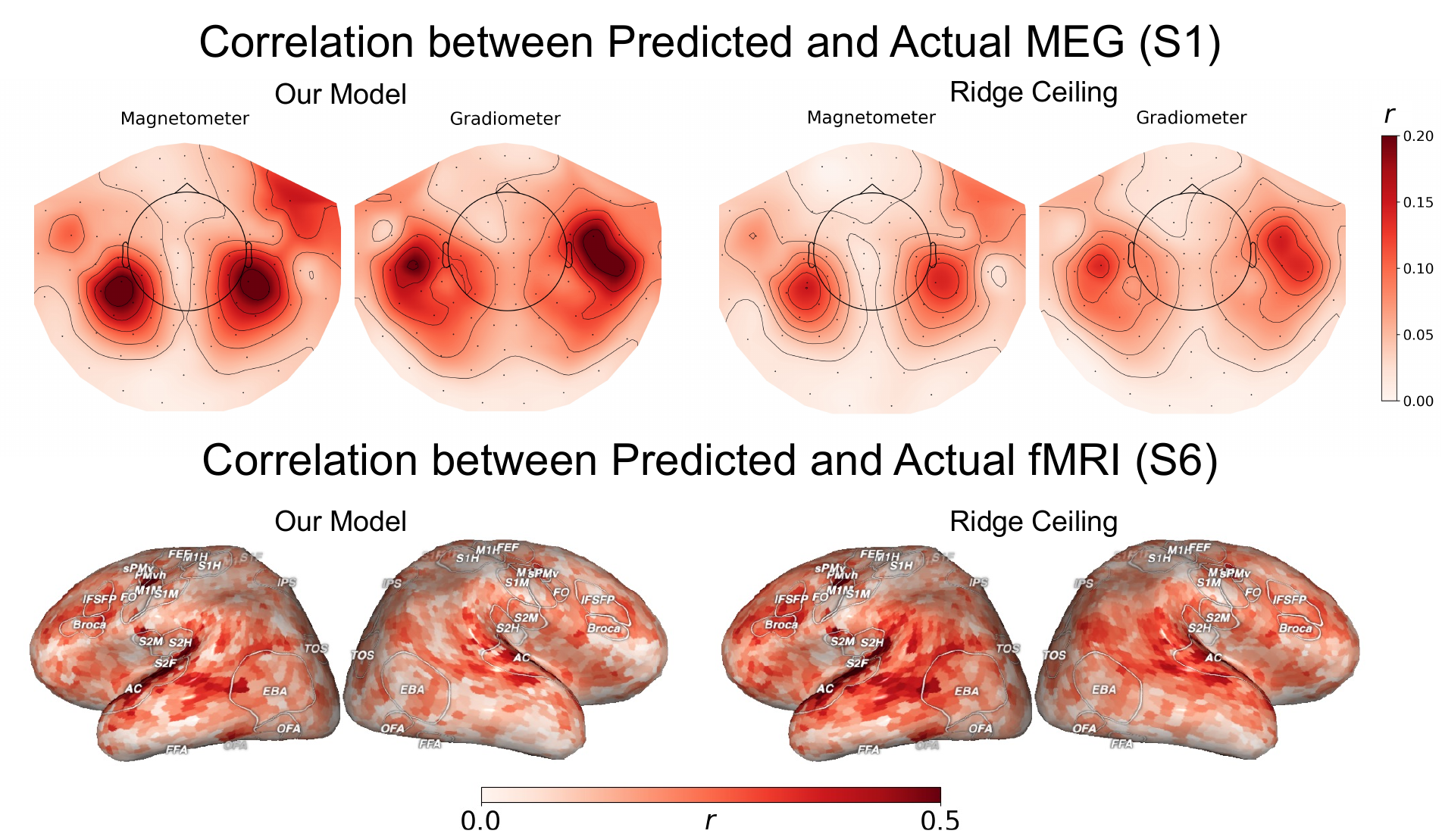}
  \vspace{-5pt}
  \caption{\textbf{Predictive performance on MEG and fMRI for two example subjects.}  Our model is comparable to single-subject, single-modality ridge models which serve as a ceiling. Top: performance on the MEG of S1. Both magnetometer and gradiometer sensors above the temporal lobe are predicted.  Bottom: performance on the source-level BOLD signals of S6, shown on the inflated surface. Large parts of bilateral temporal and frontal areas are predicted.}
  \label{fig:model_prediction}
\end{figure}

We evaluate our model and the two Ridge Ceiling models by calculating the Pearson correlation $r$ between predicted and actual signals on held-out test stories for each subject. For MEG, we report $r$ per sensor; for fMRI, we report $r$ per source. The results show that our model is comparable to both Ridge Ceiling models (Figure~\ref{fig:model_prediction} shows S1 and S6 for example). For S1's MEG, over temporal lobe sensors, the MEG Ridge Ceiling attains $r = 0.074 \pm 0.041$ whereas our model reaches $r = 0.109 \pm 0.064$. For S6's fMRI, the fMRI Ridge Ceiling yields $r = 0.267 \pm 0.074$ across the top quartile of sources, while our model achieves $r = 0.236 \pm 0.072$. The results for other subjects are reported in the appendix. We find that our model performs better than the MEG Ridge Ceiling, probably because the transformer encoder and the positional embeddings added to the keys enable our model to better capture nonlinearities and feature-dependent neural latency. However, our model performs slightly worse than the fMRI Ridge Ceiling, which might due to the fact that our model forces all subjects to share a common ``fsaverage'' source estimates and does not have a mechanism to allow for spatial variance across subjects. Crucially, unavailable from either Ridge Ceiling models, our model delivers cortical source estimates with high spatial and temporal resolution.

\section{Simulation Experiments}
\label{sec:simu}

To test the fidelity of our source estimates under controlled conditions, we use the story feature streams to generate synthetic source activity from a linear model whose weights are randomly sampled according to the empirical feature covariance, with different lags applied to word embeddings, phonemes, and spectrograms to mimic hierarchical processing in the brain. From the synthetic source activity, we then generate MEG and fMRI signals at different noise levels measured by contrast-to-noise ratios (CNR): $\infty$ (noiseless) / 1 / 0.1 for MEG and $\infty$ (noiseless) / 0.3 for fMRI\footnote{We define CNR as the ratio of the standard deviation of signal and noise: $\text{CNR}=\sigma_{\text{signal}} / \sigma_{\text{noise}}$ \citep{WelvaertRosseel_2013}. We choose these CNR levels because the mean CNR in our MEG data is around 0.1 and the mean CNR in our fMRI data is around 0.3.}.

\begin{figure}
  \centering
  \includegraphics[width=0.9\linewidth]{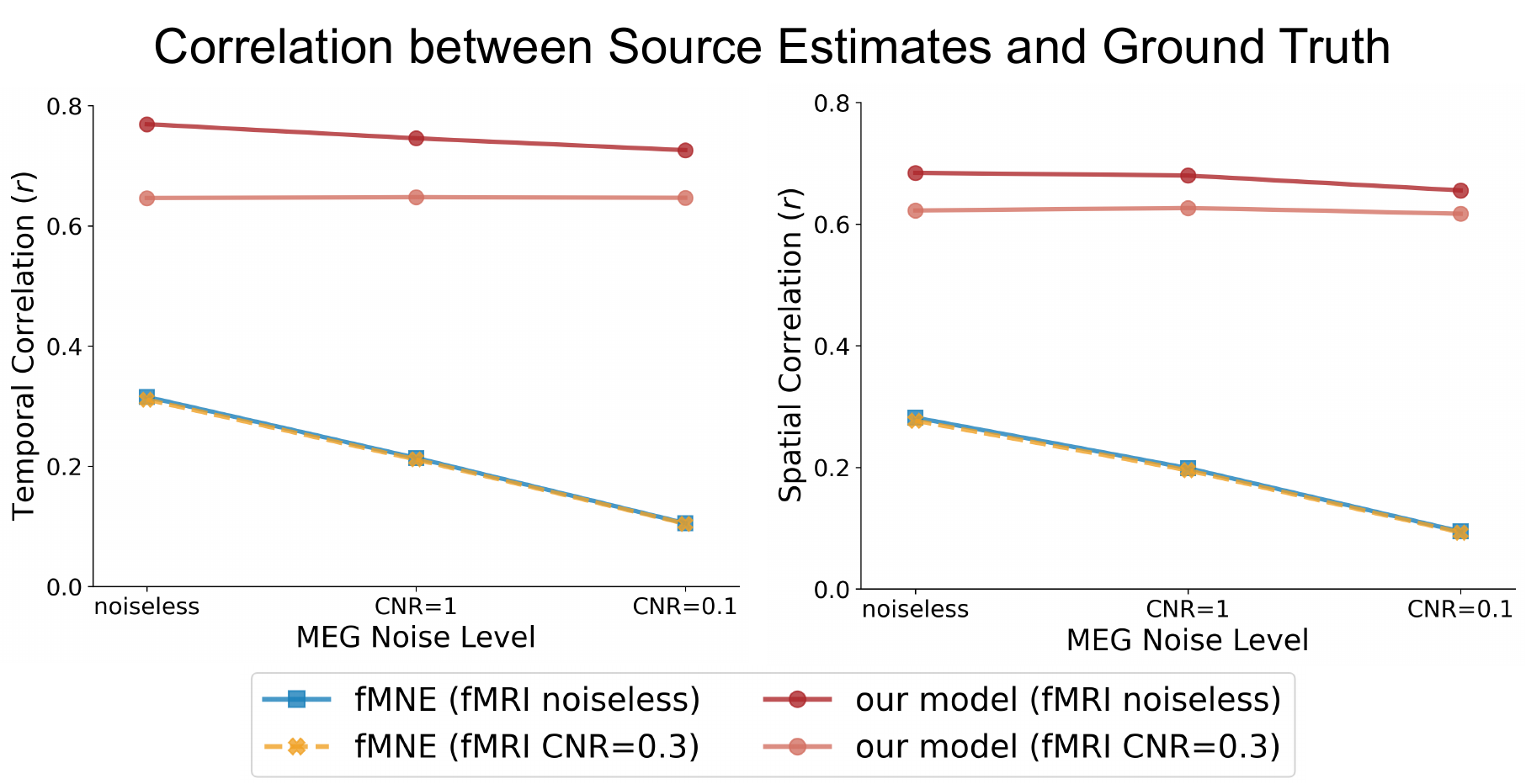}
  \vspace{-5pt}
  \caption{\textbf{Results of simulation experiments.} We report the mean Pearson $r$ computed over time within each source (left panel, temporal correlation) and over sources at each time point (right panel, spatial correlation) between the source estimates and the ground truth. Our model outperforms fMNE in both aspects under all noise levels.}
  \label{fig:simulation}
\end{figure}

We train our model on the simulated MEG and fMRI of the training stories, and evaluate the source estimates of the test stories. We calculate the Pearson $r$ between the source estimates with the ground truth along the temporal dimension or the spatial dimension. We compare our model against fMRI-weighted MNE (fMNE) \citep{Liu_1998}, which is built upon classical MNE framework \citep{Hamalainen_1994} and incorporates fMRI activity to allow sources with higher BOLD responses to be more active. Figure~\ref{fig:simulation} shows that our model recovers source activity more accurately than fMNE at all noise levels in both time and space. 

\section{ECoG Prediction}
\label{sec:ecog}

To validate the generalizability of our model, we test its predictive performance on a novel ``Podcast'' ECoG dataset \citep{Zada_2025}, which features unseen subjects and a different neural recording modality. ECoG, which comes from a small contact in the brain, provides signals with high spatiotemporal resolution, serving as a valuable proxy for ground-truth source activity.  In this dataset, neural activity was recorded from nine subjects via intracranial electrodes as they listened to a 30-minute audio podcast. Following the methodology described in Section~\ref{sec:model}, we extract semantic, phoneme, and spectrogram features from the audio stimulus. These features are then input into our trained model to generate source estimates in the ``fsaverage'' source space. 

\begin{figure}
  \centering
  \includegraphics[width=\linewidth]{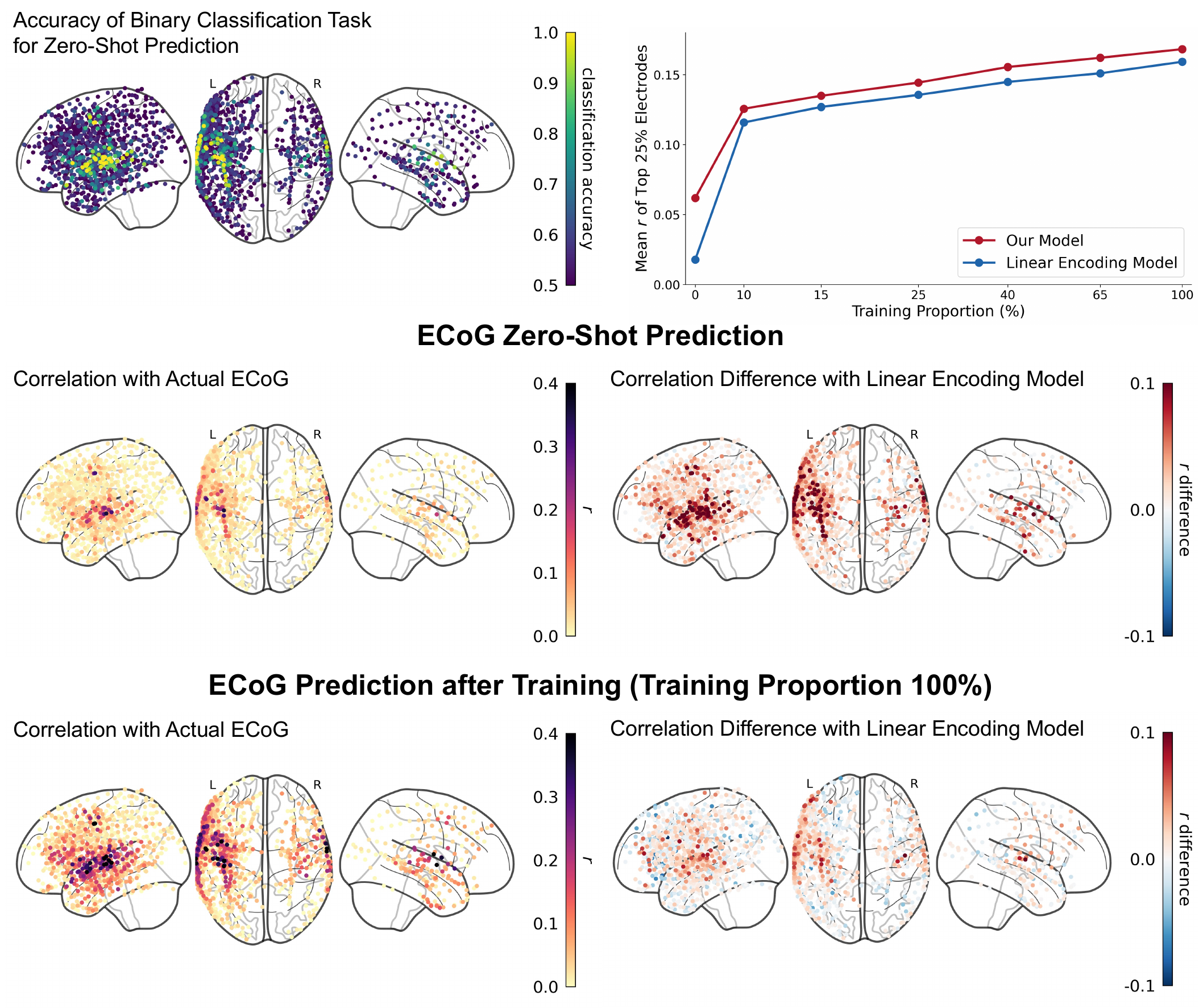}
  \vspace{-20pt}
  \caption{\textbf{Predictive performance on a new ECoG dataset.} Top left: Performance of our model's zero-shot prediction on a binary classification task. Top right: Mean Pearson $r$ of top 25\% electrodes of our model and the linear encoding model under different amount of training data. Notably, training proportion of 0\% corresponds to zero-shot prediction. Middle: Correlation map of our model's zero-shot prediction and electrode-wise correlation difference with the linear encoding model. Bottom: Correlation map of our model's trained prediction with 100\% training data and electrode-wise correlation difference with the linear encoding model.}
  \label{fig:ecog}
\end{figure}

\paragraph{Zero-Shot Prediction for ECoG}
First, we test the model's ability to generate zero-shot predictions for ECoG signals. To achieve this, we map the electrode locations to the ``fsaverage'' surface and assign the time series of the nearest estimated source as the zero-shot prediction of that electrode. We then evaluate these predictions with two analyses. First, we calculate the Pearson $r$ between each predicted and actual electrode time series, with statistical significance assessed via a permutation test. The results reveal that 916 of 1268 electrodes showed a significant correlation, with the most accurately predicted electrodes located over the superior temporal sulcus (Figure~\ref{fig:ecog}, top left panel). Second, we design a binary classification task. For each electrode, we randomly select a one-minute prediction segment and calculate its absolute Pearson $r$ with the true corresponding ECoG segment as well as with a distractor (a non-corresponding segment from the same electrode). We then measure how accurately the true segment could be identified by its higher correlation. Repeating this process 1000 times, we use the binomial test to determine significance. This evaluation finds that 683 of 1268 electrodes perform significantly above chance (Figure~\ref{fig:ecog}, middle left panel). 

\paragraph{Trained Prediction for ECoG}
In addition to zero-shot predictions, we also train a linear mapping from the source estimates to the ECoG signals. Using a three-fold cross-validation scheme, we set aside 10 minutes of ECoG data as the test set in each fold. We then train a ridge regression model to predict each electrode from our source estimates using a varying proportion of the remaining data for training. We compare the performance of this approach against a linear encoding model for ECoG trained directly on stimulus features using stacked ridge regression \citep{LinEtAl_2024}. As illustrated in Figure~\ref{fig:ecog}, when provided with an equal amount of training data, the predictions derived from our model's source estimates consistently outperform the linear encoding model. This suggests that our model provides a powerful inductive bias, generating representations that can be more readily mapped to neural activity. Notably, the performance at a training proportion of zero percent corresponds to the zero-shot prediction described in the last paragraph. In this scenario, our model's advantage is most pronounced, substantially outperforming the randomly initialized linear encoding model. Collectively, these findings demonstrate that our model's source estimates generalize effectively to new subjects and modalities, faithfully representing underlying cortical activity.

\section{Related Work}
\label{sec:relwork}

\paragraph{Source Localization}
MNE \citep{Hamalainen_1994}, beamformers \citep{Van_1997}, and Bayesian approaches \citep{Wipf_2009} are classic source localization methods for MEG/EEG. To improve spatial precision, a common strategy is to incorporate fMRI data, using activation maps as spatial priors to bias source estimates toward hemodynamically active regions \citep{Liu_1998, Dale_2000, Liu_2008, XuEtAl_2018, Henson_2010, Suzuki_2021, Moradi_2024}. However, these methods are typically applied to each time point separately and they do not leverage information from the sensory input by design, making them less suitable for tracking the rich neural dynamics elicited by naturalistic stimuli.

\paragraph{Multimodal Fusion}
Instead of using fMRI to inform MEG/EEG inversion, multimodal fusion methods use symmetric models to jointly assess information from both modalities. For example, joint independent component analysis and joint tensor/matrix decomposition on MEG and fMRI could identify shared latent spatiotemporal components \citep{Calhoun_2006,Belyaeva_2024}. Another popular approach is to align MEG and fMRI using representation similarity analysis, so that researchers can pinpoint MEG within a time window to particular cortical regions \citep{Cichy_2014,Cichy_2020,Leonardelli_2022,Yeh_2024}. However, these methods do not yield a direct, high-resolution estimate of the underlying neural source activity itself, nor do they explicitly model how that activity is driven by stimulus features.

\paragraph{Neurogenerative Modeling}
Conceptually similar to our approach, neurogenerative modeling builds models for latent neural sources and uses biophysical forward models to generate neuroimaging data \citep{Huster_2012,Castaldo_2023, Kang_2024}. For example, The Virtual Brain specifies parameterized neural mass models for cortical regions, obtains connectivity between regions from anatomical scans, uses lead-field matrices and hemodynamic functions to generate MEG/EEG and fMRI data, and fits parameters to observed data \citep{Ritter_2013,Patow_2024,Hashemi_2025}. Key distinctions from our work are that these frameworks often rely heavily on prior knowledge about neural circuits, require complicated Bayesian inference for parameter estimation, and have primarily focused on resting-state dynamics rather than stimulus encoding.

\paragraph{Encoding Models for Language}
Voxel-wise or channel-wise linear encoding with single word or contextual embeddings has mapped semantic and syntactic processing in fMRI \citep{Wehbe_2014b, HuthEtAl_2016,TonevaWehbe_2019, Schrimpf_2021, Reddy_2021, Toneva_2022, Caucheteux_2022, TangEtAl_2023} and MEG \citep{Wehbe_2014, TonevaWehbe_2019, Toneva_2022, Caucheteux_2022}.  These methods have mapped between embeddings and brain activity, pinpointing in the brain that are predicted by the information in the embeddings, and time points in which they are predicted. While MEG results have allowed researchers to paint somewhat of a spatiotemporal picture, it remains limited in the spatial resolution that it offers. The fMRI results, though with a high spatial resolution, still remain unconnected to the temporal course.  Even those works that have used both fMRI and MEG have not combined them beyond comparing their results \citep[e.g.,][]{Caucheteux_2022}.

\section{Conclusions, Limitations, and Future Work}
\label{sec:conclude}

We have presented a transformer-based encoding model that successfully recovers source activity from naturalistic, multi-subject MEG and fMRI recordings. Our model demonstrates high predictive accuracy for held-out data, and its source estimates with high spatiotemporal resolution are validated through simulation experiments and ECoG prediction. By combining MEG and fMRI with a naturalistic encoding model, our work opens new avenues for non‑invasively probing the dynamics of language and cognition without sacrificing spatial or temporal fidelity.

Our work has two limitations that point to future research. First, the current model relies on raw phoneme and spectrum features. While effective, incorporating contextualized representations from pretrained large audio-language models (e.g., Whisper \citep{Radford_2023}) could potentially improve the model's performance. Second, our source space is constrained to surface dipoles with fixed orientations. This simplification prevents the model from capturing activity subcortical structures or from non-perpendicular cortical sources. Future work should incorporate volumetric or hybrid surface–volume source spaces with orientation freedom to improve neuroanatomical fidelity.

\newpage

\bibliographystyle{plainnat}
\bibliography{references}

\newpage

\appendix

\section{Predictive Performance for All Subjects}

\begin{figure}
  \centering
  \includegraphics[width=\linewidth]{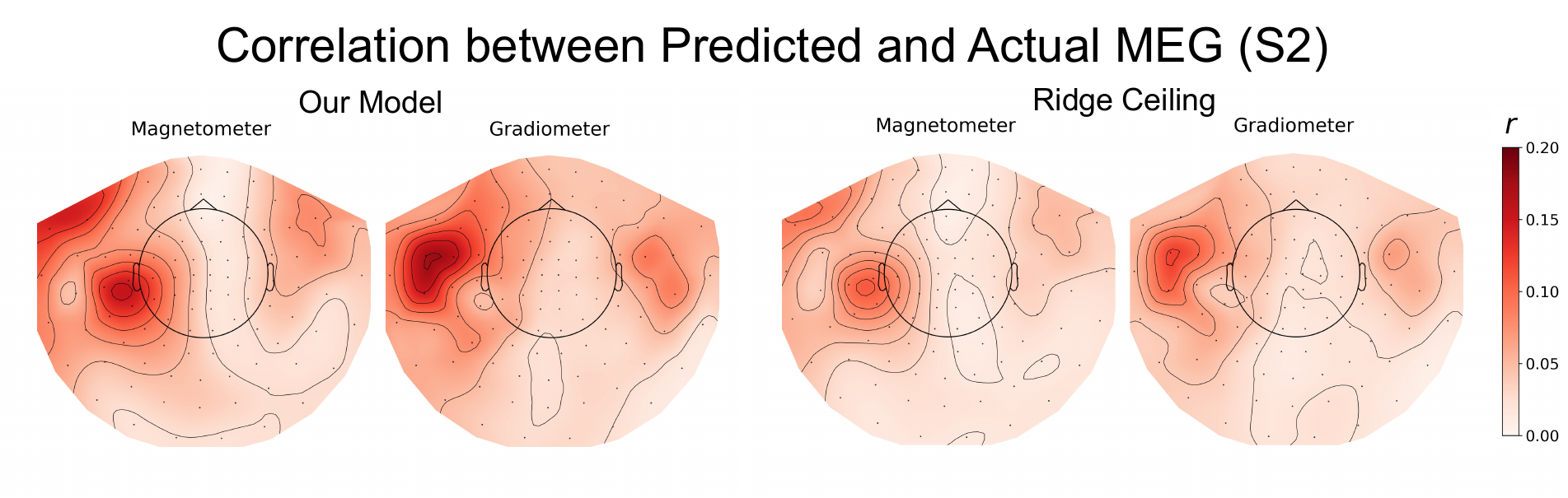} \\
  \includegraphics[width=\linewidth]{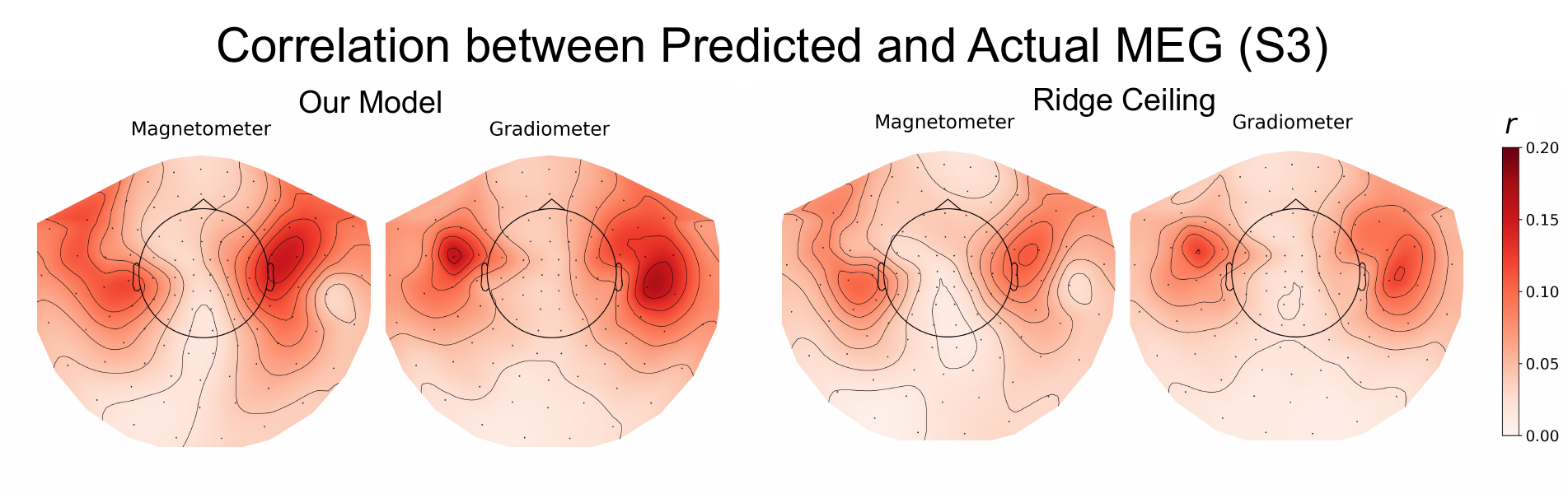} \\
  \includegraphics[width=\linewidth]{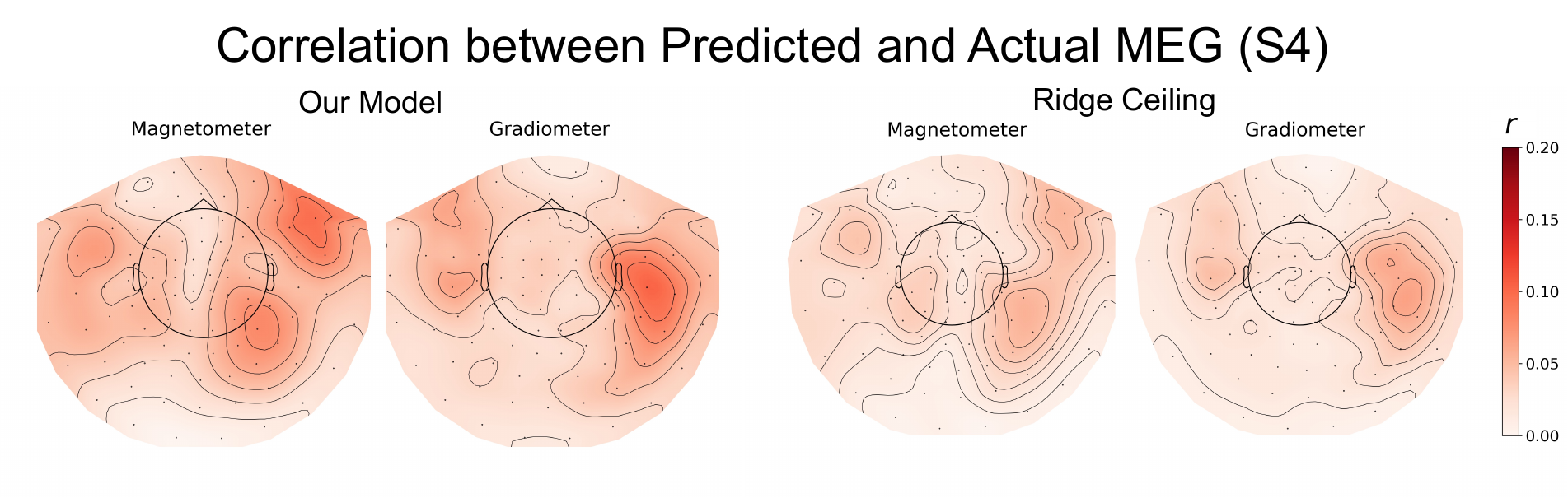} \\
  \includegraphics[width=\linewidth]{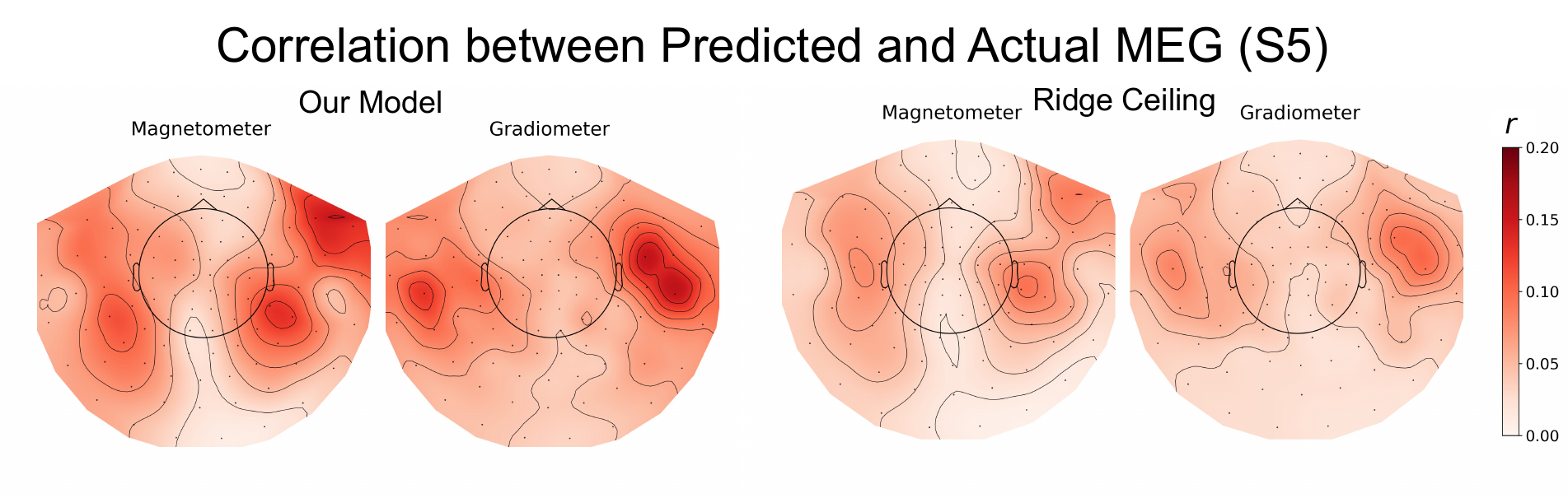}
  \caption{\textbf{Predictive performance on MEG for S2--S5.}}
  \label{fig:model_prediction2}
\end{figure}

\begin{figure}
  \centering
  \includegraphics[width=\linewidth]{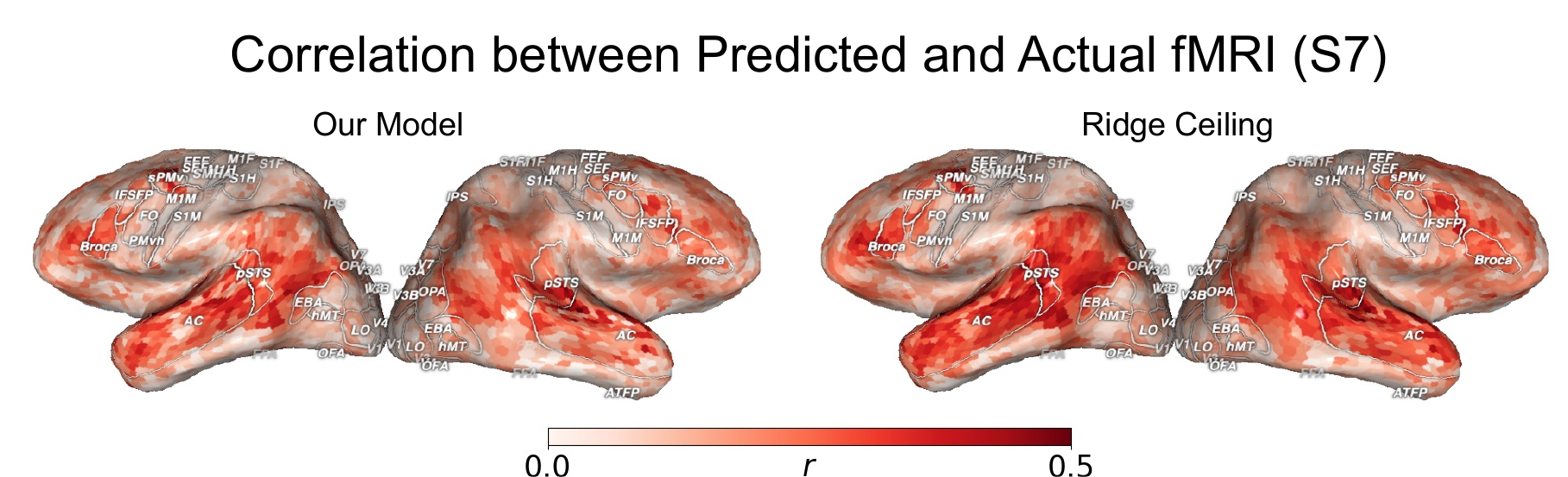} \\
  \includegraphics[width=\linewidth]{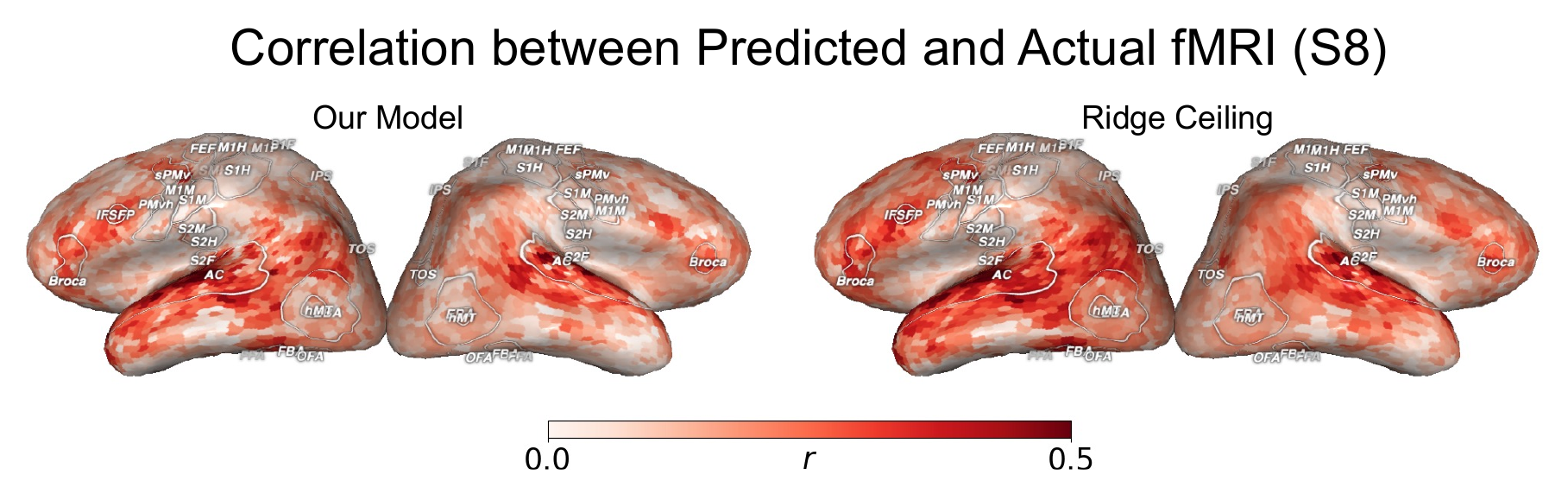} \\
  \includegraphics[width=\linewidth]{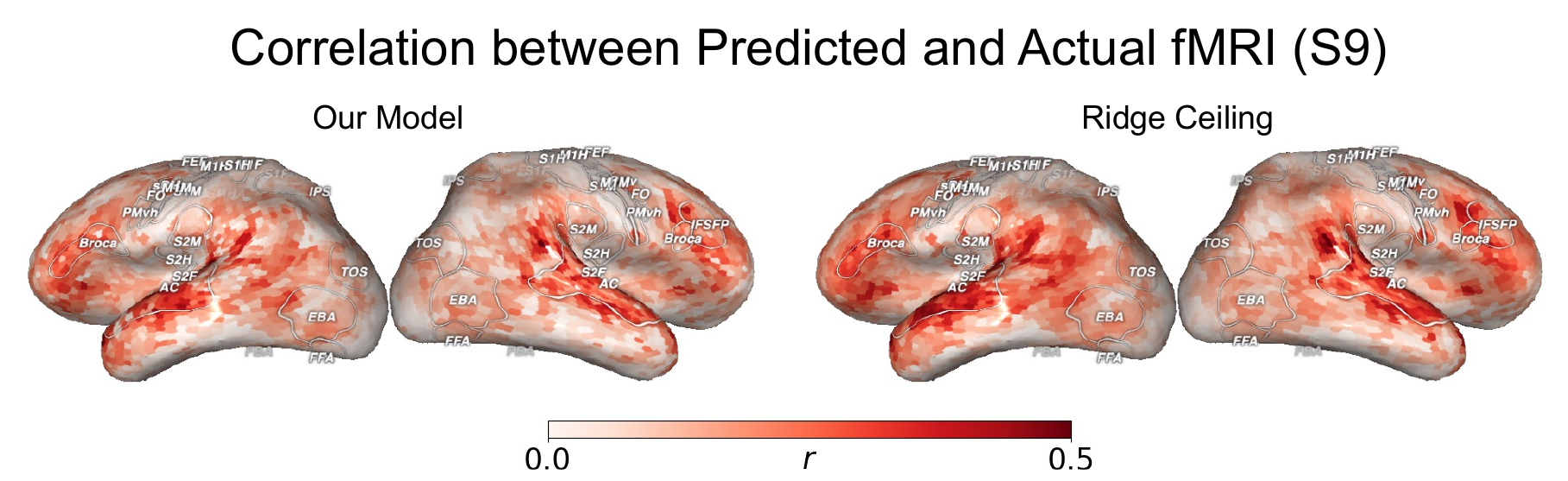} \\
  \includegraphics[width=\linewidth]{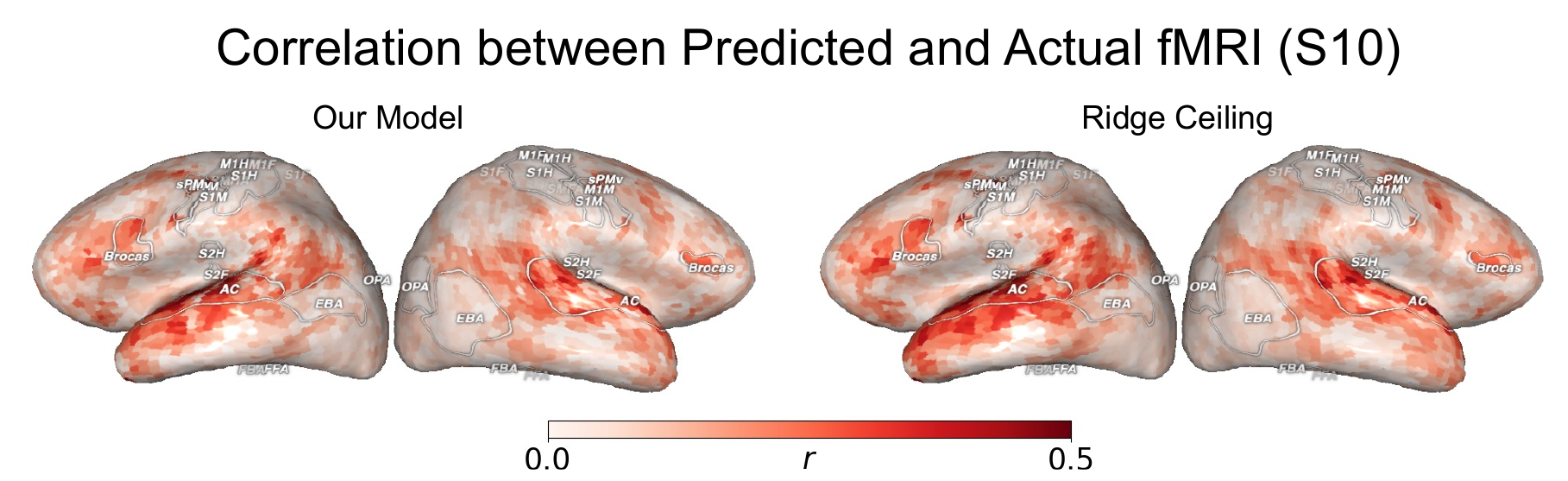}
  \caption{\textbf{Predictive performance on fMRI for S7--S10.}}
  \label{fig:model_prediction3}
\end{figure}

\begin{figure}
  \centering
  \includegraphics[width=\linewidth]{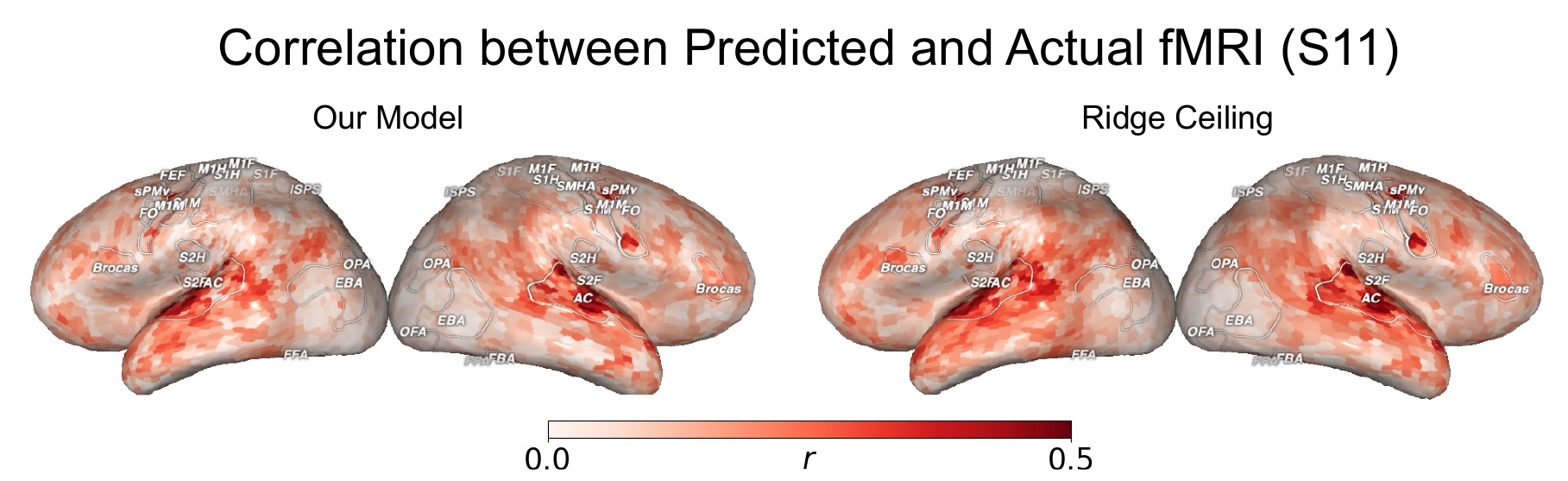} \\
  \includegraphics[width=\linewidth]{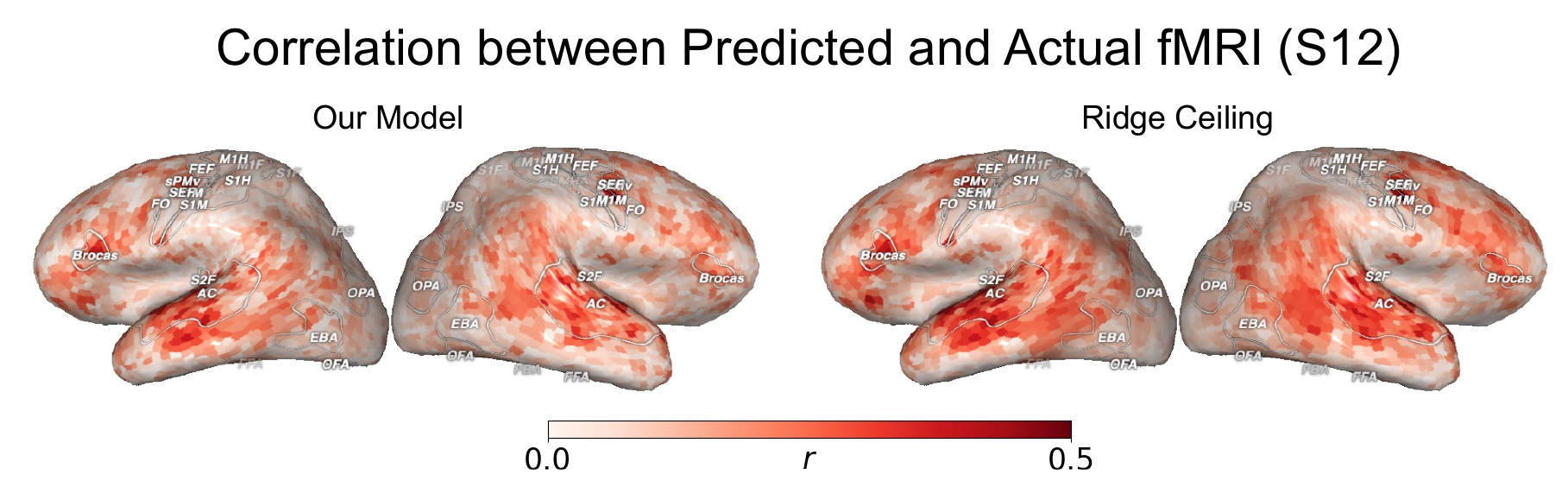} \\
  \includegraphics[width=\linewidth]{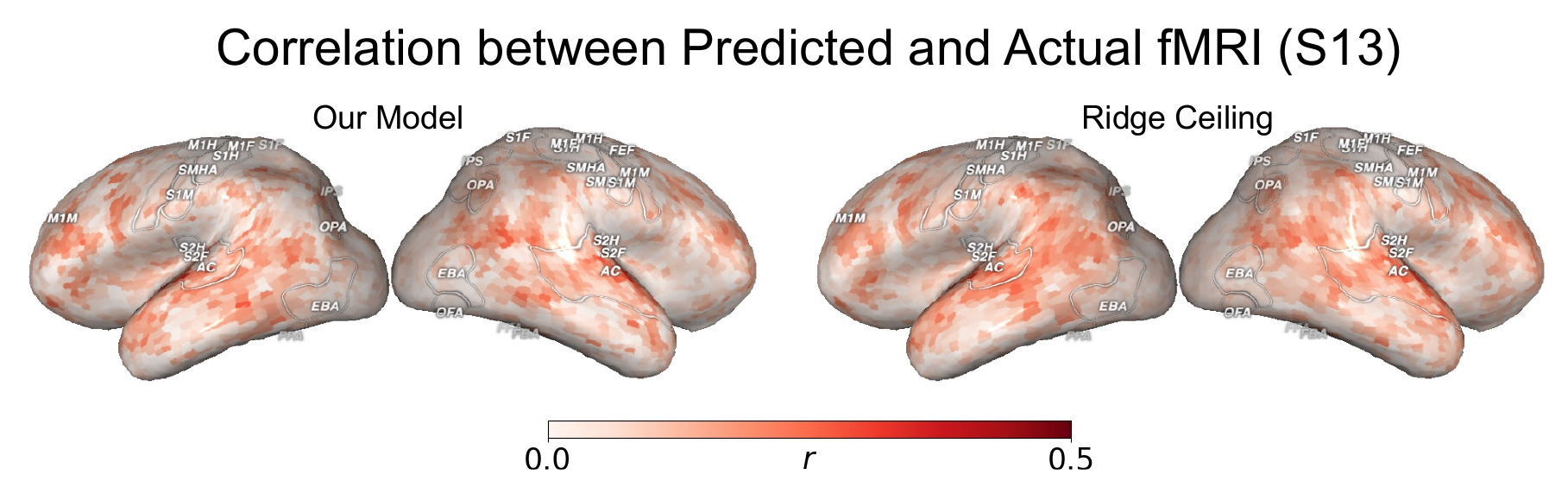}
  \caption{\textbf{Predictive performance on fMRI for S11--S13.}}
  \label{fig:model_prediction4}
\end{figure}

Here we report our model's predictive performance for MEG and fMRI for all subjects, against single-subject, single-modality Ridge Ceiling models (see Table~\ref{tab:meg}, \ref{tab:fmri}, Figure~\ref{fig:model_prediction}, \ref{fig:model_prediction2}, 
\ref{fig:model_prediction3}, \ref{fig:model_prediction4}).

\begin{table}[H]
  \caption{\textbf{MEG predictive performance} (Pearson $r$ over temporal lobe sensors)}
  \label{tab:meg}
  \centering
  \begin{tabular}{lcc}
    \toprule
    Subject     & Our Model     & MEG Ridge Ceiling \\
    \midrule
    S1 & $0.109 \pm 0.064$   & $0.074 \pm 0.041$ \\
    S2 & $0.073 \pm 0.043$   & $0.050 \pm 0.029$ \\
    S3 & $0.089 \pm 0.037$   & $0.067 \pm 0.027$ \\
    S4 & $0.054 \pm 0.024$   & $0.033 \pm 0.017$ \\
    S5 & $0.088 \pm 0.033$   & $0.059 \pm 0.021$ \\
    \bottomrule
  \end{tabular}
\end{table}

\begin{table}[H]
  \caption{\textbf{fMRI predictive performance} (Pearson $r$ of top quartile of sources)}
  \label{tab:fmri}
  \centering
  \begin{tabular}{lcc}
    \toprule
    Subject     & Our Model     & fMRI Ridge Ceiling \\
    \midrule
    S6  & $0.236 \pm 0.072$   & $0.267 \pm 0.074$ \\
    S7  & $0.234 \pm 0.056$   & $0.268 \pm 0.058$ \\
    S8  & $0.224 \pm 0.067$   & $0.262 \pm 0.064$ \\
    S9  & $0.208 \pm 0.063$   & $0.235 \pm 0.070$ \\
    S10 & $0.162 \pm 0.063$   & $0.180 \pm 0.067$ \\
    S11 & $0.180 \pm 0.062$   & $0.199 \pm 0.066$ \\
    S12 & $0.181 \pm 0.055$   & $0.210 \pm 0.059$ \\
    S13 & $0.130 \pm 0.037$   & $0.139 \pm 0.036$ \\
    \bottomrule
  \end{tabular}
\end{table}

\section{Hyperparameter Tuning}

We perform hyperparameter tuning and compare with our original design which uses four transformer layers and two attention heads for each layer. As shown in the table, in the validation set, our four-layer model outperforms the three-layer model and the five-layer model. Similarly, two attention heads yield the lowest validation loss, compared with one head or four heads.

Although these results support our chosen architecture, we wish to note that the optimal hyperparameters will likely vary depending on factors such as dataset size and stimulus complexity, and we encourage researchers to perform their own tuning when adapting our framework for specific applications.

\begin{table}[H]
  \caption{\textbf{Results of hyperparameter tuning}}
  \label{tab:hyperparam}
  \centering
  \begin{tabular}{lccc}
    \toprule
    Model & Total Validation Loss & MEG Validation Loss & fMRI Validation Loss \\
    \midrule
    \textbf{4 layers, 2 heads (current)} & \textbf{1.807} & \textbf{0.918} & \textbf{0.889} \\
    3 layers, 2 heads & 1.819 & 0.929 & 0.890 \\
    5 layers, 2 heads & 1.813 & 0.921 & 0.893 \\
    4 layers, 1 head  & 1.820 & 0.928 & 0.892 \\
    4 layers, 4 heads & 1.812 & 0.921 & 0.891 \\
    \bottomrule
  \end{tabular}
\end{table}

\section{fMNE Implementation}

We implement fMNE using \texttt{MNE-Python~1.8}. We make the inverse operator using the default settings for fixed orientation, depth weighted sources (\texttt{loose} = 0, \texttt{depth} = 0.8). We use the ground truth noise to compute the noise covariance for each MEG sensor. We set the regularization parameter \texttt{lambda2} to the inverse of the squared MEG CNR. We set the source covariance matrix as a diagonal matrix whose elements are the BOLD signal of each source at the corresponding TR after normalization.

\end{document}